\documentclass[a4paper,11pt]{article}
\usepackage{graphicx}
\usepackage{epstopdf}
\usepackage{amsfonts,amsmath,amssymb}
\usepackage{hyperref}
\usepackage{float}

\usepackage{epsfig,multicol,bbm}

\newcommand{\dbar} {\ensuremath{\,\mathchar'26\mkern-12mu d}}
\newcommand\fverb{\setbox\pippobox=\hbox\bgroup\verb}
\newcommand\fverbit{\egroup\item[\fbox{\unhbox\pippobox}]}

\newbox\pippobox

\begin{document}
\title{\bf Exact Anisotropic Solutions of the Generalized TOV Equation}
\author{Nematollah Riazi\thanks{Electronic address: n\_riazi@sbu.ac.ir},  S. Sedigheh Hashemi, S. Naseh Sajadi and  Shahrokh Assyyaee
\\
\small Department of Physics, Shahid Beheshti University, G.C., Evin, Tehran 19839,  Iran}
\maketitle
\begin{abstract}
We explore gravitating relativistic spheres composed of an anisotropic, barotropic fluid. We  assume a bi-polytropic equation of state which has a linear and a power-law terms. The generalized Tolman-Oppenheimer-Volkoff (TOV) equation which describes the hydrostatic equilibrium  is obtained. The full system of equations are solved for solutions which are regular at the origin and  asymptotically flat. Conditions for the appearance of horizon
and a basic treatment of stability are also discussed.
\end{abstract}

\section{Introduction}
Though very pioneering, the Lane-Emden equation is still a versatile and useful modeling device in astrophysics \cite{0}.
To obtain the Lane-Emden equation one needs to invoke three
essential basics: spherical symmetry,
hydrostatics equilibrium, and a polytropic equation of state \cite{10}.
 Of course, the relativistic version of the Lane-Emden
equation is now at our disposal \cite{2}. In general
relativity one needs to replace the classical hydrostatic
equilibrium equation with its relativistic dual. The
well-known TOV (Tolman-Oppenheimer-Volkoff) equation is in fact the relativistic version of
hydrostatic equilibrium \cite{3,4}\cite{11,12}.
Generally,  the Lane-Emden equation can be derived
using the polytropic equation of state in  the form $P=K\rho^{1+{1\over n}}$ with $n$ being called the polytropic index.
The rationale for postulating such a  relation between density and
pressure originates from thermodynamics. In thermodynamics, a polytropic  equation comes from the adiabatic condition
$\dbar Q=0$ where $\dbar Q$ is the transferred heat.
 One of the
immediate consequences then is $P\rho^\gamma=constant$ in which $P$ is the pressure, $\rho$ is the density and
$\gamma=\frac{C_p}{C_v}$, the ratio of the two specific heats \cite{5}. The
isothermal and the adiabatic processes are the most important
examples of a polytropic process. To find a comprehensive analysis
about the non-relativistic version of the Lane-Emden equation and its solutions
one can refer to Clayton (1968) \cite{6}.

Assuming a polytropic star makes
the relativistic structure analysis easier and more interpretable.  Solutions should
satisfy appropriate  boundary condition(s) in order to  explain various astrophysical
objects  from compact stars to galactic  objects
\cite{1}. Almost all  polytropic models assume an isotropic pressure. Here,  we allow
anisotropic cases by assuming different tangential and radial
pressure components in the energy-momentum tensor.
 All principal pressures (stresses) are assumed to have double-polytropic equations of state \cite{13}.
 A number of authors have investigated  anisotropic models (see, \cite{14}-\cite{19}).
 Assuming anisotropic stresses is by no means un-natural.
 For example, in a compact star, although the radial
pressure  vanishes at the surface, one still could
postulate a tangential pressure to exist. While the latter
does not alter the spherical symmetry, it may create some
streaming fluid motions. The Reissner-Nordstrom
solution is another relativistic solution which is supported by anisotropic stresses.
A charged black hole readily results in negative radial pressure,
while the tangential pressure remains positive. On galactic scales,
a spherical galaxy is well described by the King-Michie family of models
\cite{7,8}, which involves an anisotropic pressure. This anisotropic pressure originates from the
fact the phase space distribution function of the stars depends on coordinates and velocities
only through the constants of motion (energy and angular momentum). If the distribution function depends on the magnitude of the angular momentum, this will lead to an anisotropic pressure, even if the system is spherically symmetric.

In what follows, we first
derive the TOV equation for an anisotropic fluid with spherical symmetry (next section).  In
section \ref{3}, we will derive analytical solutions with regular behavior at the center. Section \ref{kilsec}
is devoted to a study of Killing horizons and energy conditions. The last section reflects concluding remarks.

\section{TOV equation}\label{tovsec}
Although for  most cases the classical
hydrostatic equation suffices, when the gravitational field
becomes strong,  using a general relativistic
version of the hydrostatic equilibrium equation becomes inevitable. If one starts
from a spherically symmetric metric assuming an isotropic
energy-momentum tensor  one ends with the famous TOV
equation \cite{9}
\begin{equation}\label{TOV}
-{{dP}\over {dr}}=\frac{G(\rho c^2+P)(M(r)+4\pi r^3
P)}{r^2(c^2-2GM(r)/r)}.
\end{equation}
where $P$ is the pressure and $M(r)$ is the mass variable
defined by
\begin{equation}
M(r)= \int_0 ^r{ \rho(r\rq{}) 4\pi {r^\prime} ^2 \mathrm{d}r^\prime}.
\end{equation}
From now on, we release the assumption of isotropic pressure and  let the tangential and radial pressures to be
 different. Let us take the following static and  spherically symmetric metric  for describing a gravitating relativistic sphere in Schwarzchild coordinates: $x^{\mu}=(t,r,\theta,\phi)$
\begin{equation}\label{eq1}
{\rm d}s^2= (-1+f(r))c^2{\rm d}t^2+(1-f(r))^{-1}{\rm d}r^2+r^2{\rm d}\Omega^2,
\end{equation}
 where $f(r)$ is a function of the radial coordinate $r$.
Introducing an orthonormal spherically symmetric basis
  \begin{equation}
\theta^0=(-1+f(r))^{1/2}cdt\quad ,{\theta^1}=(1-f(r))^{-1/2}dr,\quad
\theta^2=rd\vartheta,\quad\theta^3=r\sin\vartheta\,d\phi,
\end{equation}
and using $ d\theta^\mu=-\omega^\mu\,_\nu\wedge\theta^\nu$ for the
torsion-free spacetime, the nonzero components of the connection form will be
\begin{equation}\label{eq18}
\omega^0\,_1=\omega^1\,_0={{-f^\prime
(r)}\over{2(1-f(r))^{1/2}}}\theta^0,
\end{equation}
\begin{equation}
\omega^2\,_1=-\omega^1\,_2={{(1-f(r))^{1/2}}\over{r}}\theta^2,
\end{equation}
\begin{equation}
\omega^3\,_1=-\omega^1\,_3={{(1-f(r))^{1/2}}\over{r}}\theta^3,
\end{equation}
\begin{equation}\label{eq21}
\omega^3\,_2=-\omega^2\,_3={{\cot\vartheta}\over r}\theta^3.
\end{equation}
We also have the energy-momentum conservation law in the form
\begin{equation}\label{Cons}
D* T_\alpha=0,
\end{equation}
where $D$ stands for absolute exterior differential \cite{20} and $T_{\alpha}$ is the
energy-momentum 1-form
\begin{equation}
T_\alpha=T_{\alpha \beta}\theta^\beta .
\end{equation}
In our case, $*T_\alpha$ can be written in the form
\begin{equation}\label{TT}
*T^\alpha=q^\alpha \eta ^\alpha.
\end{equation}
Note that  there is no summation over indices  in (\ref{TT}), and $q^\alpha$ and
$\eta^\alpha$ are defined as
\begin{equation}\label{12}
q^\alpha=\textrm{diag}(-\rho c^2,P_r,P_t,P_t),
\end{equation}
and
\begin{equation}
\eta^\alpha=*\theta^\alpha.
\end{equation}
where
$P_{r}$ and $P_{t}$ are the radial and tangential pressures (stresses).

From the conservation law (\ref{Cons}) one obtains
\begin{equation}
dq^\alpha\wedge\eta^\alpha=-\sum_\beta\omega^\alpha\,_\beta\wedge\eta^\beta(q^\beta-q^\alpha),
\end{equation}
for $\alpha=1$ we have
\begin{equation}
dP_r\wedge\eta^1=\omega^1\,_0\wedge\eta^0(q^0-q^1)
+\omega^1\,_2\wedge\eta^2(q^0-q^2)
+\omega^1\,_3\wedge\eta^3(q^0-q^3).
\end{equation}
Substituting for the connection forms from
(\ref{eq18})-(\ref{eq21}) we obtain
\begin{align}\label{ml}
(1-f(r))^{1/2}P^\prime(r)\theta^1\wedge\eta^1=&
{{-f^\prime(r)}\over{2(1-f(r))^{1/2}}}\theta^0\wedge\eta^0(\rho c^2+P_r(r))\nonumber\\
& -{{(1-f(r))^{1/2}}\over r}(\rho
c^2-P_t(r))(\theta^2\wedge\eta^2+\theta^3\wedge\eta^3).
\end{align}
Using the metric (\ref{eq1})  and matter field (\ref{12}),  the first  two components of the Einstein field equations reduce to
\begin{equation}\label{eq5}
G^0 _{~0}= G^1_{~1}=-{1\over r} f^\prime(r)-{1\over r^2}f(r),
\end{equation}
which is equivalent to $P_{r}=-\rho c^2$.
On the other hand (\ref{eq5}) can be written as
\begin{equation}
(rf(r))^\prime=\frac{8\pi G}{c^2} \rho  r^2,
\end{equation}
where prime stands for differentiation with respect to $r$. Solving
for $f(r)$ yields
\begin{equation}
f(r)=\frac{b}{r}+\frac{2 }{rc^2} \int{4\pi G\rho {r^\prime}
^2\mathrm{d}r^\prime},
\end{equation}
with $b$ is the constant of integration. Though having a singularity
(black hole) at the origin is among the possibilities and should
be explored separately, for the moment we ignore the first term of
the above equation in the following relations. Consequently  $f(r)$ reduces to
\begin{equation}
  f(r)=\frac{2GM(r)}{c^2r},
  \end{equation}
where $M(r)$ is the gravitational mass inside  radius $r$ in such a way that
\begin{equation}\label{eq7}
M(r)=\int _{0} ^{r} 4\pi r\rq{}^2 \rho dr\rq{}.
  \end{equation}
 Substituting
for $f(r)$ and $f^\prime(r)$ in (\ref{ml}) we finally obtain
\begin{equation}\label{OTOV}
-{{dP_r}\over {dr}}=\frac{G(\rho c^2 +P_r)(M(r)c^2+4\pi r^3
P_r)}{r^2(c^2-2GM(r)/{r)}}+\frac{2(P_r-P_t)}{r},
\end{equation}
which is the generalized Tolman-Oppenheimer-Volkoff (TOV) equation. It can be seen that  Eq.  (\ref{OTOV}) readily reduces to the standard TOV
(\ref{TOV}) when  $P_r=P_t$. We will look for solutions of the generalized TOV equation which are
asymptotically Schwarzschild, in the following section.

 \section{\textbf{Analytical solutions}}\label{3}

In this section, we will derive analytical solutions of the generalized TOV equation which are non-singular throughout the space and asymptotically approach the Schwarzschild spacetime. In what follows, we assume that the radial and tangential pressures obey the following barotropic equations of state (EoS):
\begin{equation}\label{23}
P_r=w\rho c^2+w\rq{} \frac{\rho^n}{ {\rho_c}^{n-1}} c^2,
\end{equation}
\begin{equation}
P_t=w_{1}\rho c^2+w_{2}\frac{\rho^m}{\rho _{c}^{m-1}}c^2,
\end{equation}
here $\rho_{c}$ is the  central density and $w, w\rq{}, w_1, w_2$ are the dimensionless EoS  parameters. These EoSs are composed of two terms. The first term is linear which is the most widely used one in astrophysics and cosmology. The second term allows a polytropic behavior which is familiar in astrophysical applications, leading to the Lane-Emden equation in the theory of stellar structure \cite{6}. Since from Eq. (\ref{eq5}) we have the equality $P_r=-\rho c^2$ one obtains $w=-1$ and $ w\rq{}=0$ in Eq. (\ref{23}). This value for $w$ shows that we are encountering a dark-energy-like situation. Existence of negative pressure warns us about possible violation of energy conditions. However, nowadays  energy condition violating matter has become much more popular in the literature (e.g. dark energy, wormhole theory, quantum effects). In particular, the existence of a strongly negative pressure is an essential ingredient in dark energy models. Lobo \cite{lobo} presented a generalization of the gravastar picture, by considering a matching of an interior solution governed by the dark energy equation of state $w < −1/3$, to an exterior Schwarzschild vacuum solution at a junction interface. He assumed an isotropic pressure there. The present work differs from \cite{lobo} in three respects. First, the fluid pressure is assumed to be anisotropic. Second, the EoS is not assumed to be linear, and third, we do not cut off the matter density at a certain radius, thus there is no need to apply matching conditions at the surface.

  The 22 and 33 components of the Einstein equations give
\begin{equation}\label{eq8}
G^2_{~2}=G^3_{~3}=-\frac{1}{2}f''(r)-{1\over r} f^\prime(r).
\end{equation}
Eqs.\,(\ref{eq5}) and (\ref{eq8}) can be simplified by introducing  $u \equiv {1\over r}$
in the following form
\begin{equation}\label{9}
-u^3f_u+u^2f=\frac{8\pi G }{c^2}\rho ,
\end{equation}
\begin{equation}\label{10}
u^4f_{uu}=-\frac{16\pi G}{c^4} P_t,
\end{equation}
respectively.
Here, the subscript $u$ indicates derivative with respect to the variable $u$.
Taking  derivative with respect to $u$ from Eq. (\ref{9}) and
then multiplying by $u$ and finally  using Eq. (\ref{10}), one  obtains
\begin{equation}
\rho(u)(1+w_1)+w_2\frac{\rho ^{m}(u)}{\rho_{c}^{m-1}}=\frac{1}{2}\frac{d\rho(u)}{d \ln u},
\end{equation}
which  can be solved for density $\rho(r)$:
\begin{equation}\label{eq12}
\rho(r)=\frac{\rho_c }{ [ \beta(    \frac{r}{r_{0} }     )^{2(1+w_{1})(m-1)}       -\alpha   ]^{\frac{1}{m-1}}},
\end{equation}
where
\begin{equation}\label{13}
\beta \equiv (x_{0}^{1-m} +\alpha), \quad x_0= \frac{\rho(r_{0})}{\rho_c}, \quad \alpha  \equiv \frac{w_2}{1+w_1}, \quad r_0 \equiv \frac{1}{u_0}.
\end{equation}
Eq. (\ref{eq12}) is the general form of the density in all of the  models which will be discussed separately in the following sub-sections.
At the center of the star  $r=0$ the boundary condition  $\rho(r)=\rho_c$  implies $\alpha^2=1$ corresponding to  $\alpha =\pm 1$ if $1+w_1>0$.

\subsection{Case 1}
\begin{center}
\begin{figure}[H] \hspace{4.cm}\includegraphics[width=8.cm]{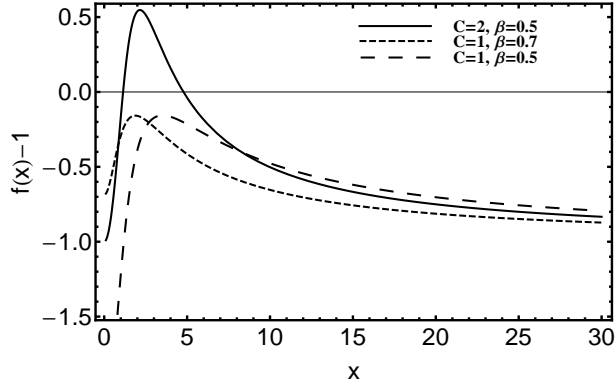}\caption{\label{Case1} \small
 The metric function  $g_{00}=f(x)-1$ as a function of $x\equiv r/r_0$ is plotted for Case $I$.  Note that there are two coordinate singularities (Killing horizons)  for the values $C=2$ and $\beta=0.5$. For  $C=1$ and $\beta=0.7$, there are no Killing horizons. For  $C=1$ and $\beta=0.5$ there are
no Killing horizons a naked singularity exists at $x=0$.}
\end{figure}
\end{center}

\begin{center}
\begin{figure}[H] \hspace{4.cm}\includegraphics[width=8.cm]{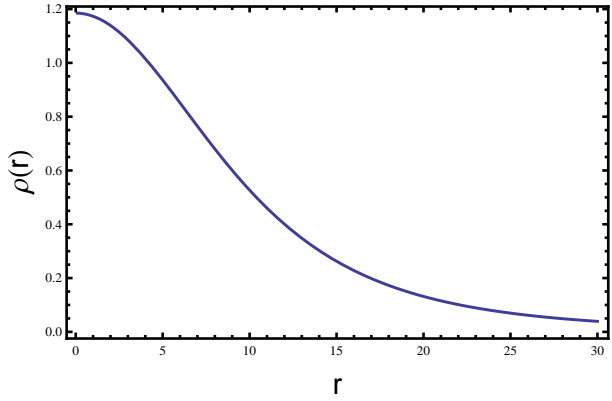}\caption{\label{ro1} \small
Energy density $\rho(r)$ as a function of radial distance $r$ is depicted for case 1.}
\end{figure}
\end{center}

   The first case corresponds to $ m=\dfrac{3}{2} $ in Eq.\ (\ref{eq12}). By  substituting the values $ \alpha=-1 $ and $ w_{2}=-2 $ into Eq. (\ref{13}),  we obtain $ w_{1}=1 $. Now the tangential pressure of the anisotropic star for this special case  becomes
\begin{equation}
 P_t=\rho c^2-2{\rho^\frac{3}{2} \over \sqrt{\rho_c}} c^2,
\end{equation}
the mass function can be obtained by performing the integration  in Eq. (\ref{eq7})
\begin{equation}
M(r)=\dfrac{2\pi\rho_{c}{r_{0}}^{3}}{\beta^{\frac{3}{2}}}\left[ \arctan(\sqrt{\beta}(\dfrac{r}{r_{0}}))-\dfrac{\dfrac{r}{r_{0}}}{2\beta(\beta\frac{r^{2}}{{r_{0}}^{2}} +1)}\right],
\end{equation}
and the function $f(r)$ becomes
\begin{equation}
f(r)=\dfrac{2GM(r)}{c^{2}r}=\dfrac{4\pi G \rho_{c}{r_{0}}^{3}}{\beta^{\frac{3}{2}}c^{2}r}\left[\arctan(\sqrt{\beta}(\dfrac{r}{r_{0}}))-\dfrac{\dfrac{r}{r_{0}}}{2\beta(\beta\frac{r^{2}}{{r_{0}}^{2}} +1)}\right].
\end{equation}
 The total gravitational mass of the system will be
\begin{equation}
M=\lim_{r\longrightarrow\infty} M(r)=\dfrac{\pi^{2}\rho_{0}{r_{0}}^{3}}{\beta^{\frac{3}{2}}},
 \end{equation}
therefore, the metric is asymptotically flat $ (f\rightarrow0) $.
Furthermore, at the center the mass function becomes zero, as expected for a non-singular model.
Thus this case satisfies the physical requirements of a  non-singular relativistic gravitating system.\\
The metric and density functions are shown in Figures \ref{Case1} and \ref{ro1}, respectively.

\subsection{Case 2}

\begin{center}
\begin{figure}[H] \hspace{4.cm}\includegraphics[width=8.cm]{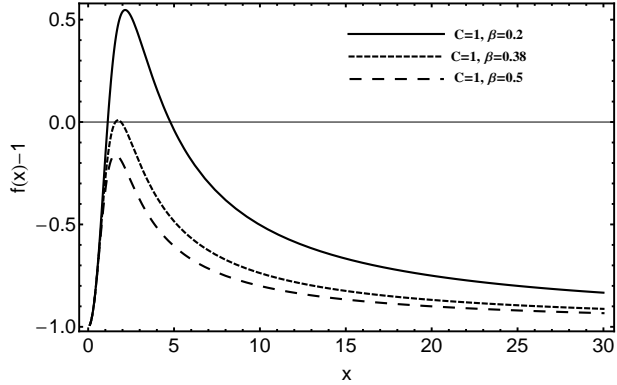}\caption{\label{Case2} \small
 The metric function $g_{00}$ as a function of the dimensionless radius for case 2. For $C=1$ and $\beta=0.2$, there are two Killing horizons. For $C=1$ and $\beta=0.38$ the two Killing horizons become degenerate.  Finally, with $C=1$ and $\beta=0.5$ there are no Killing horizons. This case does not have any curvature singularities.}
\end{figure}
\end{center}

\begin{center}
\begin{figure}[H] \hspace{4.cm}\includegraphics[width=8.cm]{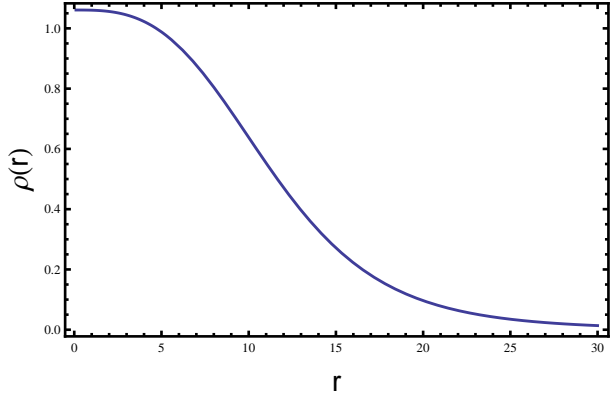}\caption{\label{ro2} \small
Energy density $\rho(r)$ as a function of radial distance $r$ is plotted for case 2.}
\end{figure}
\end{center}

Setting  $ \alpha=-1 $ and $ w_{2}=-3$ we get $ w_{1}=2 $, the tangential equation of state becomes

\begin{equation}
P_{t}=2\rho c^{2}-3\dfrac{\rho^{\frac{3}{2}}c^{2}}{\sqrt{\rho_{c}}},
\end{equation}
by substituting $ w_{2} $ in Eq. (\ref{13}), the mass function is obtained as
\begin{equation}
M(r)=\dfrac{4\pi{\rho_{c}}{r_{0}}^{3}}{3}\dfrac{(\frac{r}{r_{0}})^{3}}{1+\beta(\frac{r}{r_{0}})^{3}},
\end{equation}
furthermore,
\begin{equation}
f(r)=\dfrac{2GM(r)}{c^{2}r}=\dfrac{8\pi G{\rho_{c}}{r_{0}}^{3}}{3c^{2}r}\dfrac{(\frac{r}{r_{0}})^{3}}{1+\beta(\frac{r}{r_{0}})^{3}}.
\end{equation}
The total gravitational mass of the sphere is
\begin{equation}
M=\lim_{r\longrightarrow\infty} M(r) =\dfrac{4\pi{\rho_{c}}{r_{0}}^{3}}{3\beta}.
\end{equation}
The metric for this case is asymptotically flat and at the center of star the mass function equals zero.
The metric and density functions  for this case are shown in Figures \ref{Case2} and \ref{ro2}, respectively.

\subsection{Case 3}

\begin{center}
\begin{figure}[H] \hspace{4.cm}\includegraphics[width=8.cm]{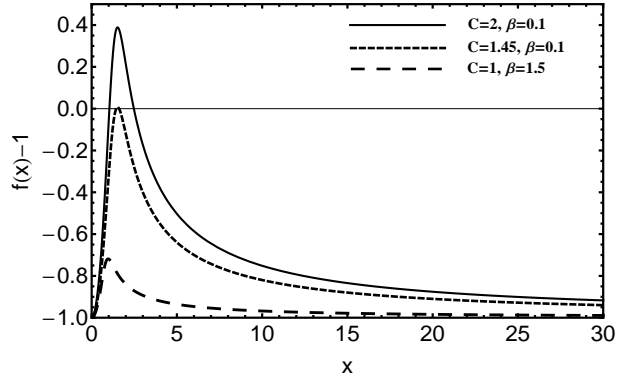}\caption{\label{Case3} \small
Two Killing horizons and a geometric singularity at $x=0$ for case 3, corresponding to $C=2$ and $\beta=0.1$. $C=1.45$ and $\beta=0.1$ leads to degenerate Killing horizons. For $C=1, \beta =1.5$, there are no Killing horizons but a naked singularity exists at $x=0$.}
\end{figure}
\end{center}

\begin{center}
\begin{figure}[H] \hspace{4.cm}\includegraphics[width=8.cm]{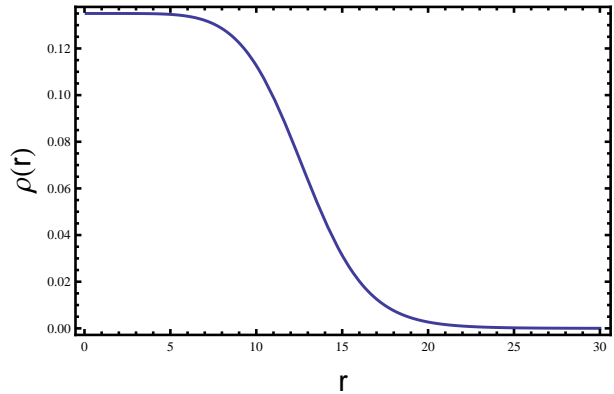}\caption{\label{ro3} \small
Energy density $\rho(r)$ as a function of radial distance $r$ for case 3.}
\end{figure}
\end{center}

Setting  $ \alpha=-1 $ and $ w_{2}=-6$ we have $ w_{1}=5 $, and the tangential equation of state becomes
\begin{equation}
P_{t}=5\rho c^{2}-6\dfrac{\rho^{\frac{3}{2}}c^{2}}{\sqrt{\rho_{c}}},
\end{equation}
the mass function is obtained as
\begin{equation}
M(r)=\dfrac{\pi{\rho_{c}}{r_{0}}^{3}}{3}\left[ \dfrac{(\dfrac{r}{r_{0}})^{3}}{1+\beta(\dfrac{r}{r_{0}})^{6}}+\dfrac{\arctan\sqrt{\beta}(\dfrac{r}{r_{0}})^{3} }{\sqrt{\beta}}\right],
\end{equation}
and
\begin{equation}
f(r)=\dfrac{2GM(r)}{c^{2}r}=\dfrac{2\pi G{\rho_{c}}{r_{0}}^{3}}{3c^{2}r}\left[ \dfrac{(\dfrac{r}{r_{0}})^{3}}{1+\beta(\dfrac{r}{r_{0}})^{6}}+\dfrac{\arctan\sqrt{\beta}(\dfrac{r}{r_{0}})^{3} }{\sqrt{\beta}}\right].
\end{equation}
The total gravitational mass is obtained as
\begin{equation}
M=\dfrac{\pi^{2}{\rho_{c}}{r_{0}}^{3}}{6\sqrt{\beta}}.
\end{equation}
Therefore, the metric is asymptotically flat and at the center of the system the mass function vanishes.
The metric and density functions  for this case are shown in Figures \ref{Case3} and \ref{ro3}, respectively.

\subsection{Case 4}
When $ \alpha=-1 $ and $ w_{2}=-1$ then $ w_{1}=0 $ and the tangential pressures becomes

\begin{equation}
P_{t}=-\dfrac{\rho^{\frac{3}{2}}c^{2}}{\sqrt{\rho_{c}}},
\end{equation}
and the mass function will be given by
\begin{equation}
M(r)=\dfrac{4\pi{\rho_{c}}{r_{0}}^{3}}{\beta^{2}}\left[\dfrac{r}{r_{0}}-\dfrac{2\ln(\beta\dfrac{r}{r_{0}}+1)}{\beta}-\dfrac{1}{\beta(\beta\dfrac{r}{r_{0}}+1)}+\dfrac{1}{\beta}\right],
\end{equation}
and
\begin{equation}
f(r)=\dfrac{2GM(r)}{c^{2}r}=\dfrac{8\pi G{\rho_{c}}{r_{0}}^{3}}{\beta^{2}c^{2}r}\left[ \dfrac{r}{r_{0}}-\dfrac{2\ln(\beta\dfrac{r}{r_{0}}+1)}{\beta}-\dfrac{1}{\beta(\beta\dfrac{r}{r_{0}}+1)}+\dfrac{1}{\beta}\right],
\end{equation}
 However, the total mass diverges in this case. In order to have physically viable solution, the total mass should be finite. Moreover, we have a singular mass at the origin.\\

\subsection{Case 5}
When $ \alpha=-1 $ and $ w_{2}=\dfrac{3}{2}$ then $ w_{1}=\dfrac{1}{2} $ and the tangential pressure can be expressed as

\begin{equation}
P_{t}=\dfrac{\rho c^{2}}{2}-\dfrac{3\rho^{\frac{3}{2}}c^{2}}{2\sqrt{\rho_{c}}}.
\end{equation}
The mass function will be
\begin{equation}
M(r)=\dfrac{8\pi{\rho_{c}}{r_{0}}^{3}}{3\beta^{2}}\left[ \dfrac{3}{1+\beta(\dfrac{r}{r_{0}})^{\frac{3}{2}}}+\ln(1+\beta(\dfrac{r}{r_{0}})^{\frac{3}{2}})-1\right],
\end{equation}
and
\begin{equation}
f(r)=\dfrac{2GM(r)}{c^{2}r}=\dfrac{16\pi G {\rho_{c}}{r_{0}}^{3}}{3\beta^{2}c^{2}r}\left[ \dfrac{3}{1+\beta(\dfrac{r}{r_{0}})^{\frac{3}{2}}}+\ln(1+\beta(\dfrac{r}{r_{0}})^{\frac{3}{2}})-1\right].
\end{equation}
In this case, too, we do not have a regular behavior at the infinity and the center, since
\begin{equation}
r\longrightarrow \infty, \quad M\longrightarrow\infty \quad  \textrm{and} \quad  f\longrightarrow 0,
\end{equation}
\begin{equation}
r\longrightarrow 0, \quad M\longrightarrow \dfrac{16\pi{\rho_{c}}{r_{0}}^{3}}{3\beta^{2}} \quad \textrm{and}\quad f\longrightarrow \infty.
\end{equation}

We have also reached analytical solutions for $ (w_{1}=\dfrac{3}{2}, w_{2}=-\dfrac{5}{2}) $ and $ (w_{1}=3, w_{2}=-4) $. However, the solutions are too lengthy to be shown here.

\section{Killing Horizons and Energy Conditions}\label{kilsec}
Since the spacetimes considered here are all static, we have the Killing vector $K=\partial _{t}$  and
 Killing horizon for the above cases can be obtained by putting the time-time component of the metric (\ref{eq1}) equal to zero which gives $f(x_{H})=1$, where $x_{H}\equiv \frac{r_H}{r_0}$.
 Unfortunately, analytical solutions for $r_H$ could not be found.  Therefore, we demonstrate  qualitative $f(x)-1$  behavior of the function for different values of $C$ and $\beta$ by plotting  the cases mentioned in the previous section (see Figures (\ref{Case1})-(\ref{Case3}). Note that  we have defined $C\equiv \frac{8\pi G \rho_c r_0^2}{3c^2}$ for convenience.

As it is seen in Fig. (\ref{Case1}) which is plotted for  Case $1$, the function  $f(x)-1$ has two killing horizons  for the values $C=2$ and $\beta=0.5$
with a singularity at $x=0$. With  $C=1$ and $\beta=0.7$ there is no Killing horizon. For  $C=1$ and $\beta=0.5$
there are no Killing horizons  but there is a   singularity at $x=0$.
Fig. (\ref{Case2})  is plotted for Case $2$. With $C=1$ and $\beta=0.2$, there are two Killing horizons. For $C=1$ and $\beta=0.38$ a degenerate root exists which is regarded as the extremal Killing horizon. Finally, with $C=1$ and $\beta=0.5$ there is no Killing horizon. It should be noted that this case does not have any geometrical singularity.

 In Fig. (\ref{Case3}), there are two Killing horizons with a singularity at $x=0$, corresponding to $C=2$ and $\beta=0.1$ are shown. $C=1.45$ and $\beta=0.1$ leads to an extremal Killing horizon. For $C=1, \beta =1.5$ we
 do not have any Killing horizon but there is a singularity at $x=0$.

The metric function is not plotted for cases 4 and 5, since they are not physically interesting.

Here, we  examine weak energy condition $ \textbf{(WEC)} $ and strong energy conditions $ \textbf{(SEC)} $ for the cases introduced in the previous section.
 \textbf{WEC} requires $ T_{\mu\nu}V^{\mu}V^{\nu}\geq0 $ for any non-spacelike vector field $ V^{\mu} $ which leads to \cite{7}
\begin{equation}\label{eq49}
\rho\geq0,\;\;\;\;\;\rho+P_{r}\geq0,\;\;\;\;\;\rho+P_{t}\geq0.
\end{equation}
 $ \textbf{SEC} $ states that $ T_{\mu\nu}V^{\mu}V^{\nu}\geq\dfrac{1}{2}T^{\lambda}_{~\lambda}V^{\sigma}V_{\sigma} $
for any timelike vector $ V^{\mu} $, or
\begin{equation}
\rho+P_{r}\geq0,\;\;\;\rho+P_{t}\geq0,\;\;\;\rho+P_{r}+2P_t \geq0.
\end{equation}
For Case $1$, Eq. (\ref{eq49}) reduces to $\rho\geq0$  and   $ \rho\leq\rho_{c} $ which  are both satisfied since  the density is maximum at the center and the density is assumed to be positive everywhere.
Therefore, $ \textbf{WEC} $ is  fully satisfied.  \textbf{SEC} leads to $\rho_c \geq 4\rho$ which is also satisfied.
Thus this case satisfies the physical requirements of a realistic relativistic model.

 \textbf{WEC} is satisfied for Case $2$, similar to the previous case.
\textbf{SEC} is also satisfied, provided that  $ \rho_{c}\geq\dfrac{9}{4} \rho $.
  This case also satisfies the physical properties that should be satisfied by a realistic model. For the  remaining cases, the  \textbf{WEC} remains satisfied as before.   \textbf{SEC} for Case $3$, implies $\rho_c\geq\dfrac{36}{25}\rho$. However, for Case $4$ we obtain $\rho\leq 0$ which contradicts positive energy theorem.
Finally, for the last case,  \textbf{SEC} requires $\rho_c\geq9\rho$.

\section{Stability and Horizon}
As  it was shown,  there appear Killing Horizons for some cases. The surface gravity $\kappa$ associated with the  time-like Killing vector $K=\partial_{t} $ can be obtained via \cite{20}
\begin{equation}
\kappa^{2}=-\dfrac{1}{4}\left\langle dK,dK\right\rangle\vert_{H}.
\end{equation}
$\kappa$ can be calculated for all of the cases considered before. As an example, for the second case we  have
\begin{equation}
\kappa=\pm \dfrac{1}{2}\acute{f(r)}\vert_{r_{H}}=\pm \dfrac{4\pi G\rho_{c}}{3c^{2}}\left[ \frac{2r_H}{1+\beta (\dfrac{r_{H}}{r_{0}})^3}- \frac{3 \beta r_{H}^4 }{r_0^3(1+ \beta(\dfrac{r_{H}}{r_{0}})^3)^2}\right].
\end{equation}

The area of the outer Killing  horizon
 can be found straightforwardly by setting $ r=r_{H} $ and $dt=dr=0 $ in Eq. (\ref{eq1}) yielding
\begin{equation}
{\rm d}s^2= r_{H}^2{\rm d}\Omega^2.
\end{equation}
The horizon area is then the integral of the induced volume element,
\begin{equation}
A=\int \sqrt{\vert g\vert}\;{\rm d}\theta\; {\rm d}\phi,
\end{equation}
so the area of the Killing  horizon  is simply
\begin{equation}
A=4\pi r_{H}^{2}.
\end{equation}
Therefore, the area of the Killing  horizon for all cases is given by the usual black hole area.
The question whether the derived solutions are  stable or not requires a separate  full study. However, as a
first check, we have plotted the total mass versus central density for the case 1 to case 3 models. It can be seen from these figures that the total gravitational mass has a minimum at certain central densities, signalling gross stability.

\begin{center}
\begin{figure}[H] \hspace{4cm}\includegraphics[width=8.cm]{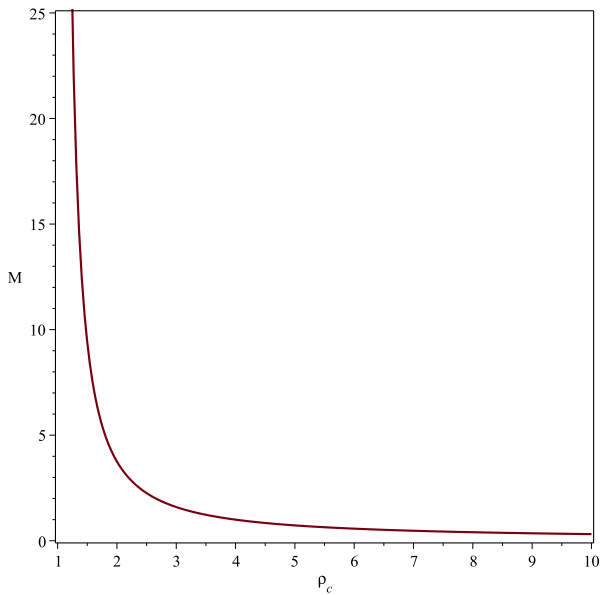}\vspace{1cm}\caption{\label{stability1} \small
 Total mass as a function of central density for case 1 model (all parameters are set to one).}
\end{figure}
\end{center}

 \begin{center}
\begin{figure}[H] \hspace{4.cm}\includegraphics[width=8.cm]{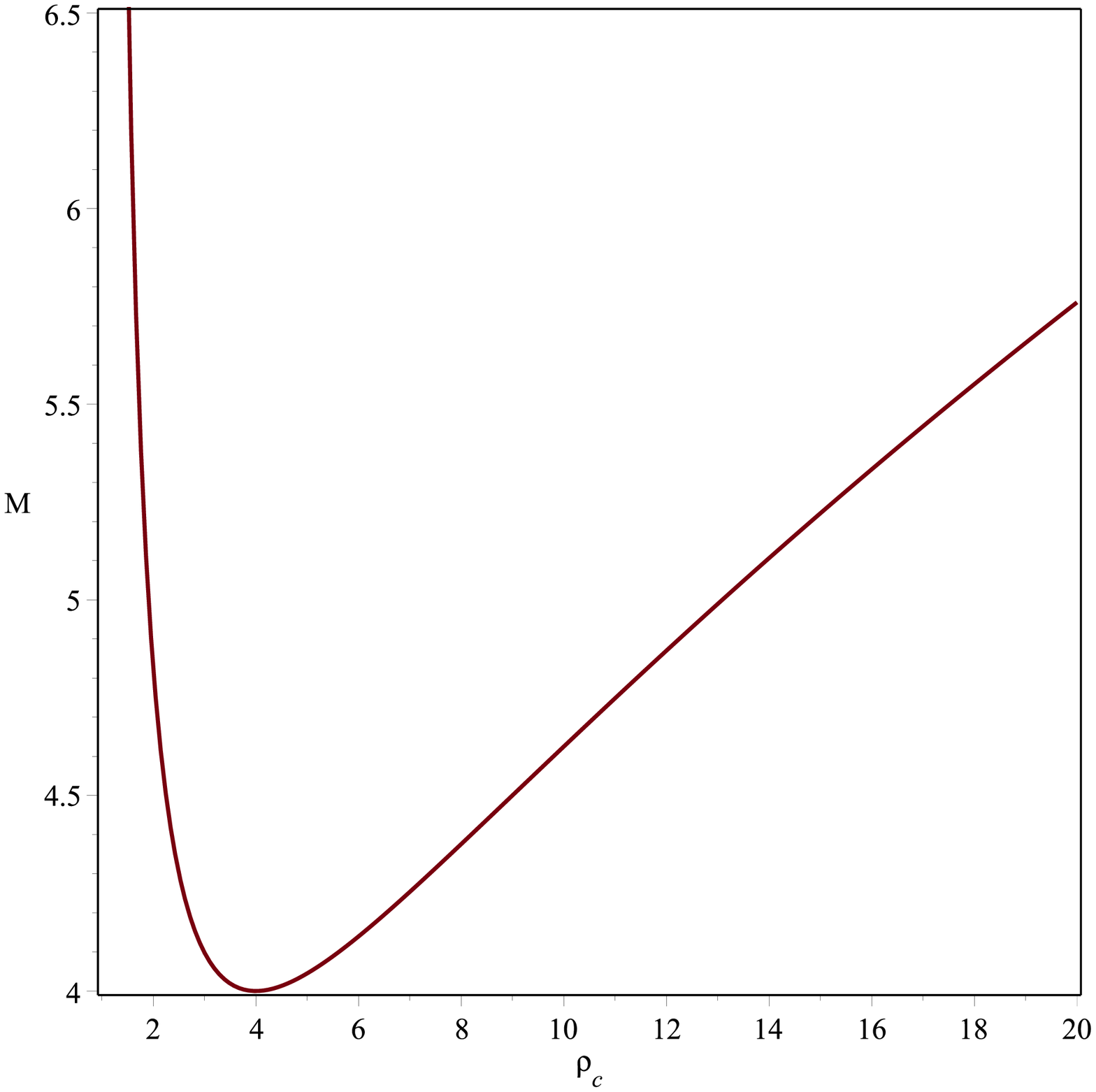}\vspace{1cm}\caption{\label{stability2} \small
 Total mass as a function of central density for case 2 model (all parameters are set to one).}
\end{figure}
\end{center}

 \begin{center}
\begin{figure}[H] \hspace{4.cm}\includegraphics[width=8.cm]{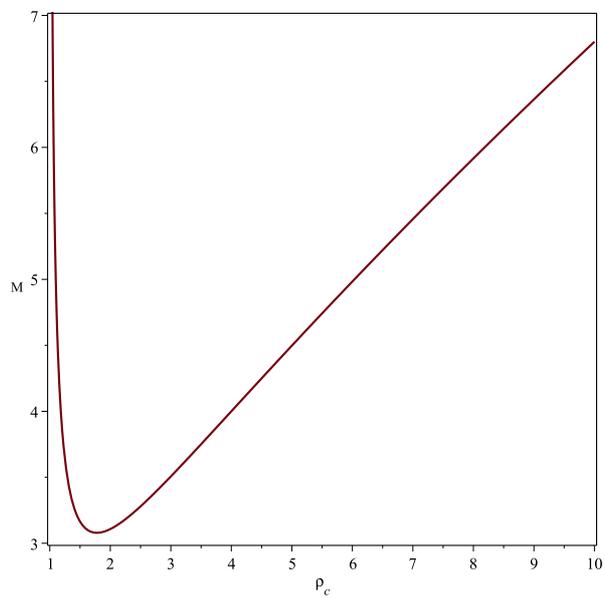}\vspace{1cm}\caption{\label{stability3} \small
 Total mass as a function of central density for case 3 model (all parameters are set to one).}
\end{figure}
\end{center}

Note that when there are two horizons, the coordinates $t$ and $r$ become spacelike and timelike, respectively, in the region between the two horizons. Thus the situation is somehow similar to the non-extremal Reissner-Nordstrom black hole.
 \section{Summary and Conclusion}
  In this paper, we started with generalizing the TOV equation for a gravitating relativistic sphere with an anisotropic, barotropic fluid. We adapted an equation of state which has a  linear term plus a power-law term which is encountered in various polytropic fluids. The full system of equations was then solved exactly for solutions which are regular at $r=0$ and smoothly approach an asymptotically flat spacetime. Some singular and asymptotically non-flat cases were also mentioned, without going into details. Finally, conditions for the emergence of horizons were examined. It was shown that in most cases, it is possible to avoid horizons, have two horizons, and also have a single degenerate horizon. The status of weak and strong energy conditions were also discussed. Finally, case 2 and case 3 models were shown to be stable, in the sense that the total gravitational mass is minimized for certain central densities.\\ \ \\

  {\bf Acknowledgements}\\

Authors acknowledge the support of Shahid Beheshti University.

\end{document}